\documentclass[12pt,epsfig]{article}
\usepackage{epsfig}
\usepackage{amsmath}
\usepackage{hhline}
\usepackage{amssymb}
\usepackage{times}
\usepackage{cite}

\newlength{\dinwidth}
\newlength{\dinmargin}
\def\herwig{{\sc Herwig} \hspace{.05cm}}
\def\pomwig{{\sc Pomwig} \hspace{.05cm}}

\def\C2q{C_2(Q)}
\def\tC2q{$\C2q$}
\def\f2q{f_2(Q)}
\def\tf2q{$\f2q$}

\newcommand{\PO}{I\!\!P}
\newcommand{\RO}{I\!\!R}

\begin{document}

\sloppy

%\begin{titlepage}

  \begin{flushright}
    MAN/HEP/2001/03 \\
    MC-TH-01/09 \\
    October 2001
  \end{flushright}
  \begin{center}
    
    \vskip 10mm {\Large\bf\boldmath Double Diffractive Higgs and Di-photon Production at the Tevatron and LHC} \vskip 15mm

    {\large Brian Cox and Jeff Forshaw}\\
    Dept.~of Physics and Astronomy, University of Manchester\\
    Manchester M13 9PL, England\\
    coxb@mail.desy.de
    \vskip 10mm

    {\large Beate Heinemann}\\
	Oliver Lodge Laboratory, University of Liverpool \\
        Liverpool, L69 7ZE, England\\
    	beate@fnal.gov	
    \vskip 10mm

  \end{center}
  \vskip 0mm
\begin{abstract}

We use the \pomwig Monte Carlo generator to predict the cross-sections for double diffractive higgs and di-photon production at the Tevatron and LHC. We find that the higgs production cross-section is too small to be observable at Tevatron energies, and even at the LHC observation would be difficult. Double diffractive di-photon production, however, should be observable within one year of Tevatron Run II.

\end{abstract}

%\end{titlepage}

\section{Introduction}

With the start of Run II at the Tevatron the possibility of detecting a light higgs boson has of course been the focus of much attention. One suggested experimentally clean search strategy is to look in double diffractive collisions in which the final state contains only intact protons, which escape the central detectors, and the decay products of the higgs, in this case two {\it b} quark jets.  The protons would be tagged in roman pot detectors a long way down stream of the interaction point, allowing a very precise determination of the higgs mass, using the so-called missing mass method \cite{albrow}. With such proton tagging, the available center of mass energy of up to 200 GeV means that the preferred higgs mass of around 115 GeV is kinematically accessible, at least in principle. The missing mass method can be used in the case where higgs production is {\it exclusive}, that is the process $p + p \rightarrow p + {\rm gap} + H + {\rm gap} + p$. Unfortunately, recent calculations of this rate indicate that it is too low to be observable at the Tevatron \cite{Khoze:2000cy}. The {\it inclusive} process, $p + p \rightarrow p + {\rm gap} + H + X + {\rm gap} + p$, however, was recently estimated by Boonekamp et al. \cite{Boonekamp:2001vk} to be large enough to be detectable. If this turns out to be the case, then the double diffractive channel might still provide a relatively clean environment in which to produce the higgs. In this paper, we use the \pomwig Monte Carlo \cite{pomwig} to predict the  inclusive rate. \pomwig implements an Ingelman-Schlein type model of diffraction \cite{IS} which has proved successful in describing a range of diffractive processes in DIS at HERA \cite{Adloff:1997sc}. We describe the \pomwig physics model in Section 2.    

We also consider two other double diffractive processes: dijet production and di-photon production. The CDF collaboration has measured double diffractive dijet production rates in Run I \cite{cdfdijet}, and we can therefore compare our predictions to these data. Ingelman-Schlein type models such as \pomwig typically predict rates higher than those observed at the Tevatron\footnote{For a recent review see \cite{Cox:2000yr} and references therein.}. Our strategy in this paper is to attribute this difference to rapidity gap survival probability: interactions between spectator partons in the beam protons can destroy a rapidity gap produced in a diffractive interaction \cite{Bj}. We assume that the factor necessary to scale our dijet predictions to data is to a good approximation independent of the hard subprocess, and use it to rescale our higgs and di-photon production cross-section predictions. 

We expect the di-photon process to be an interesting process to measure in the first years of Run II since, as we shall see, the rates are almost certainly high enough to be observed. Di-photon production is dominated by the quark-quark diagram of Figure \ref{fig:gamgam}. This is in contrast to the dijet and higgs production processes which are dominated by gluon exchange. The di-photon process therefore provides complementary information to the dijet process about the structure of the diffractive exchange in a very clean experimental environment. 
       
\section{Double diffraction in \pomwig}
\pomwig generates diffractive hard scattering events in the spirit of the Ingelman-Schlein model. That is to 
say it assumes that the production cross-section factorises into a product of regge flux factor and 
corresponding parton distribution function. At present, two regge exchanges are included which we refer
to as ``pomeron exchange'' ($\PO$) and ``reggeon exchange'' ($\RO$). In the current version of \pomwig it is not possible to mix
the two exchanges. For particular details of the \pomwig default parameterisations (which we use in this
paper) see \cite{pomwig}. 

\begin{figure}[htb]        
\begin{center}
  \setlength{\unitlength}{1 mm}      
  \Large
  \begin{picture}(140,60)(0,0)
    \put(20,0){\epsfig{file=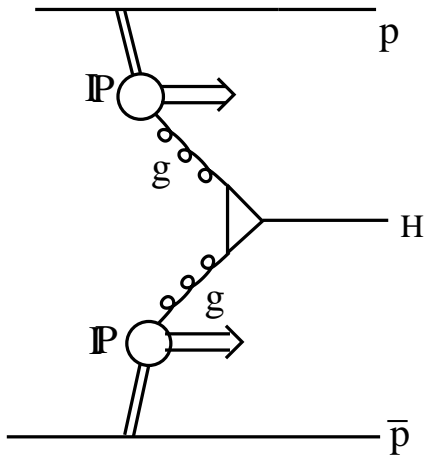,width=0.35\textwidth}}

  \end{picture}
\end{center}
\caption {\label{fig:higgs} Diffractive higgs production in \pomwig}
\end{figure}

Figure \ref{fig:higgs} illustrates the mechanism by which we generate higgs boson production in \pomwig. In all cases,
the higgs is accompanied by ``pomeron remnants''. This whole system is separated by rapidity gaps from
the outgoing hadrons due to the requirement that the energy fraction lost by the incoming hadrons, $\xi$,
is small (the transverse momentum transfer to the outgoing hadrons is also small). More specifically the total cross-section for producing higgs bosons in events
where the incoming hadrons lose a fraction of their momenta less than $\xi_{max}$ (assuming the top quark is sufficiently heavy) is 
\begin{equation}
\sigma_H \approx \frac{G_F \alpha_s^2}{288 \pi \surd 2} \tau \int_\tau^1 \frac{dx}{x} g_1(x,m_h^2)
g_2(\tau/x,m_h^2)
\end{equation}
where $\tau = m_h^2/s$ ($\surd s$ is the hadron-hadron centre-of-mass energy) and the diffractive
gluon density for hadron $i$ in pomeron exchange is
\begin{equation}
g_i(x,Q^2) = \int_x^{\xi_{max}} d\xi_i f_{\PO/i}(\xi_i) g_{\PO}(x/\xi_i,Q^2).  
\end{equation}
$f_{\PO/i}(\xi_i)$ is referred to as the pomeron flux factor and $g_{\PO}(\beta,Q^2)$ is the gluon density
of the pomeron, where $\beta$ is the fractional momentum of the pomeron carried by the gluon. Similar expressions exist for the subleading exchange which we generically refer to
as ``reggeon''. We use the diffractive parton densities as extracted from diffractive deep inelastic
scattering measurements by the H1 collaboration \cite{Adloff:1997sc}.
  
\begin{figure}[htb]        
\begin{center}
  \setlength{\unitlength}{1 mm}      
  \Large
  \begin{picture}(140,60)(0,0)
    \put(0,0){\epsfig{file=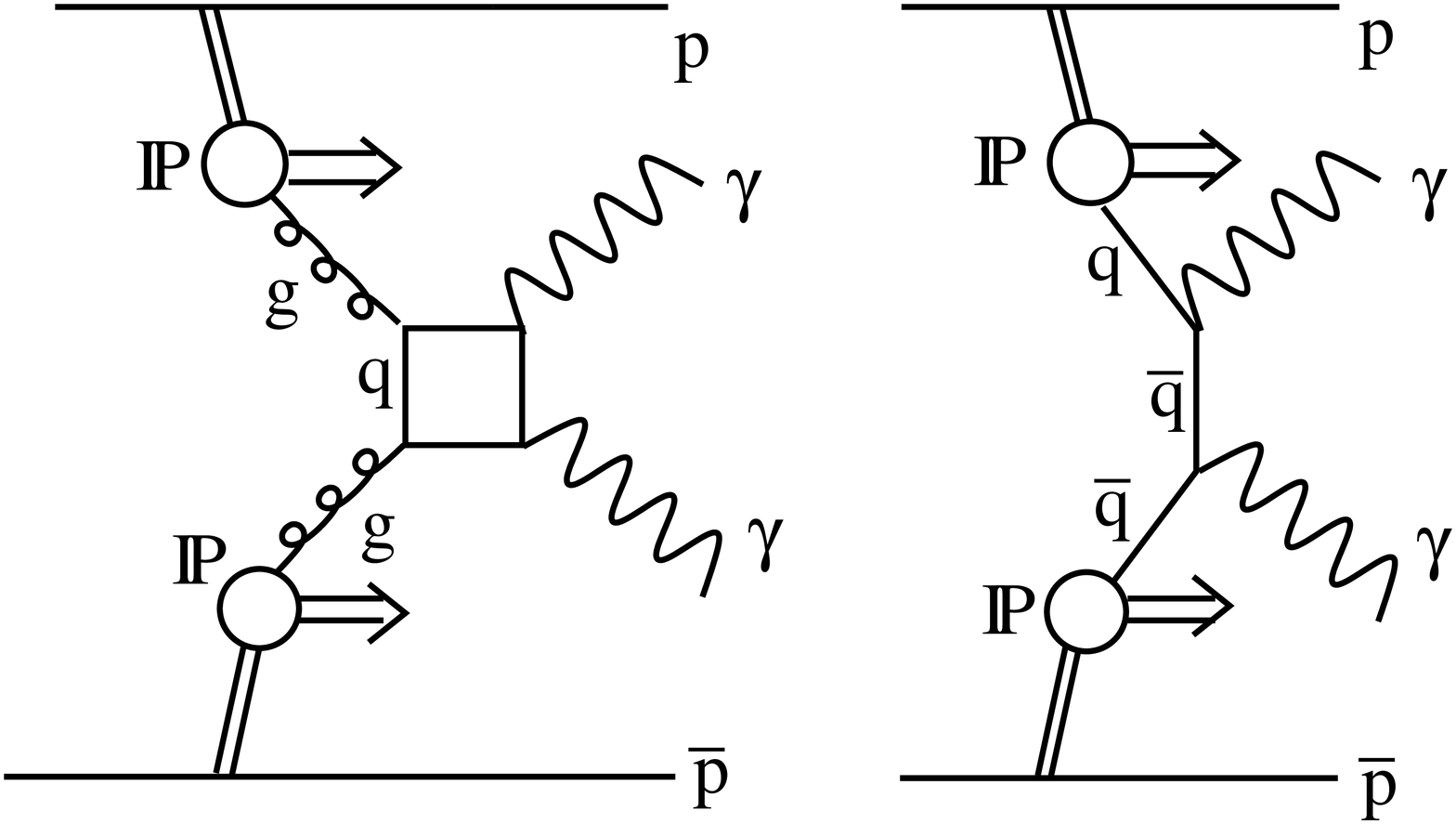,width=0.6\textwidth}}
  \end{picture}
\end{center}
\caption 
        {\label{fig:gamgam} Di-photon production in \pomwig }

\end{figure}

The relevant \pomwig subprocesses for di-photon production are shown in Figure \ref{fig:gamgam}. The hard subprocess is as in \herwig 6.1 \cite{HERWIG}.

\section{Results}
\begin{figure}
\begin{picture}(250,85)
\put(0,84){\makebox(0,0){(a)}}
\put(83,84){\makebox(0,0){(b)}}
\put(-10,0){\epsfig{file=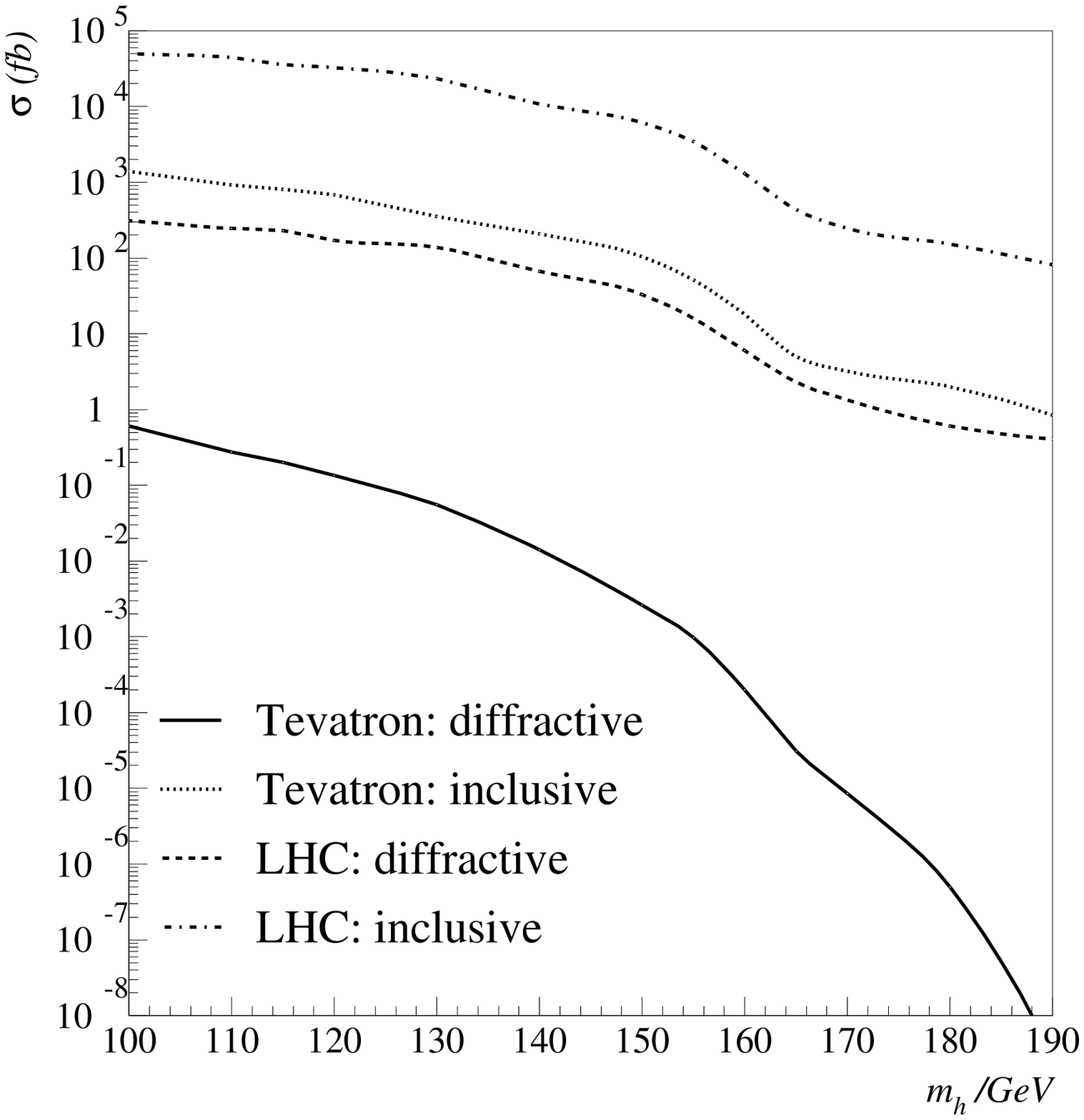,width=.55\textwidth}}
\put( 73,0){\epsfig{file=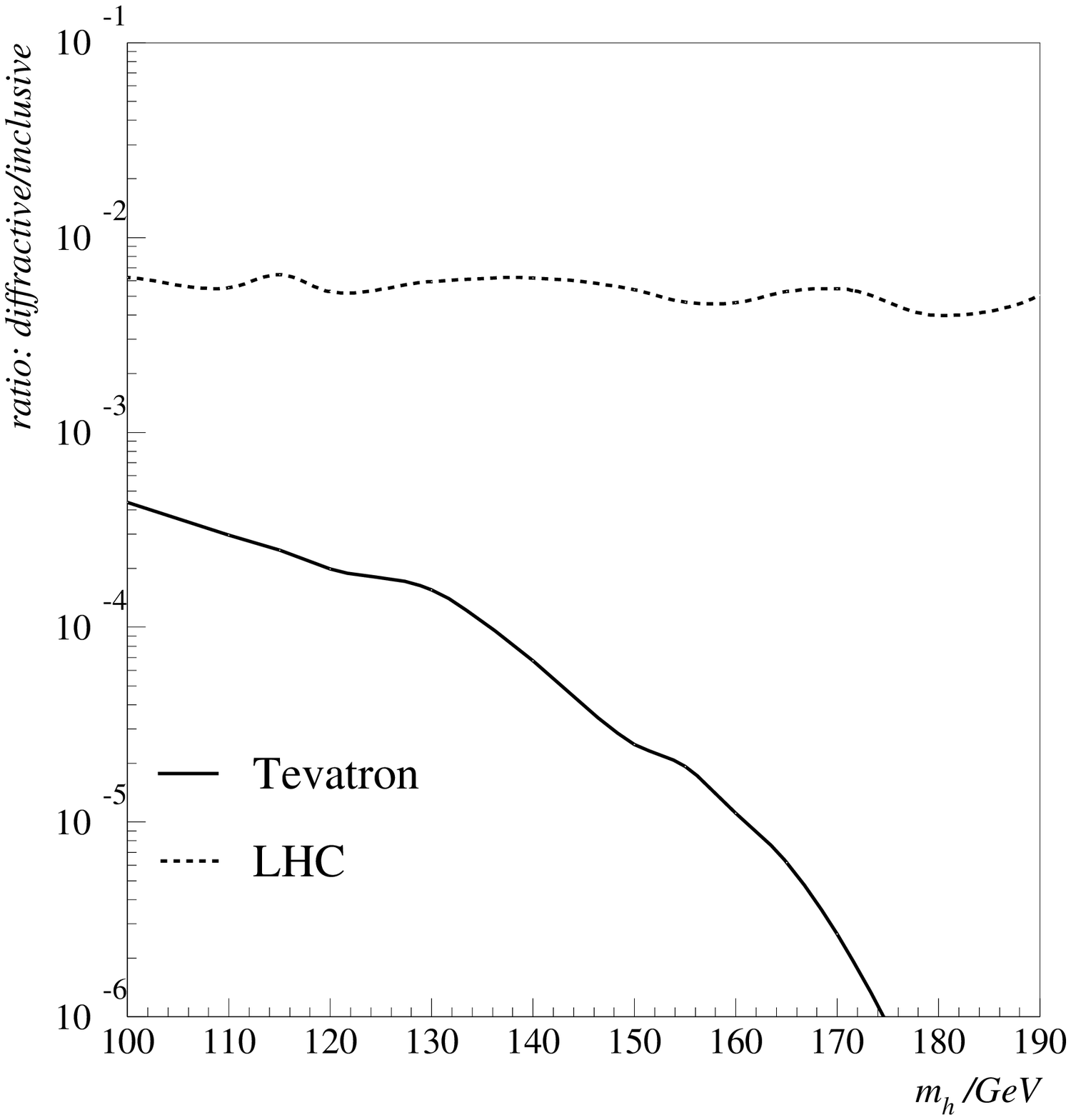,width=.55\textwidth}}
\end{picture}
\caption{The inclusive and double diffractive higgs production cross-sections at Tevatron and LHC energies as a function of higgs mass are shown in (a). In (b) the ratio of the double diffractive to inclusive higgs production cross-sections is shown.}
\label{hgcs}
\end{figure}
\begin{figure}

\begin{picture}(250,85)

\put(0,84){\makebox(0,0){(a)}}
\put(83,84){\makebox(0,0){(b)}}
\put(-10,0){\epsfig{file=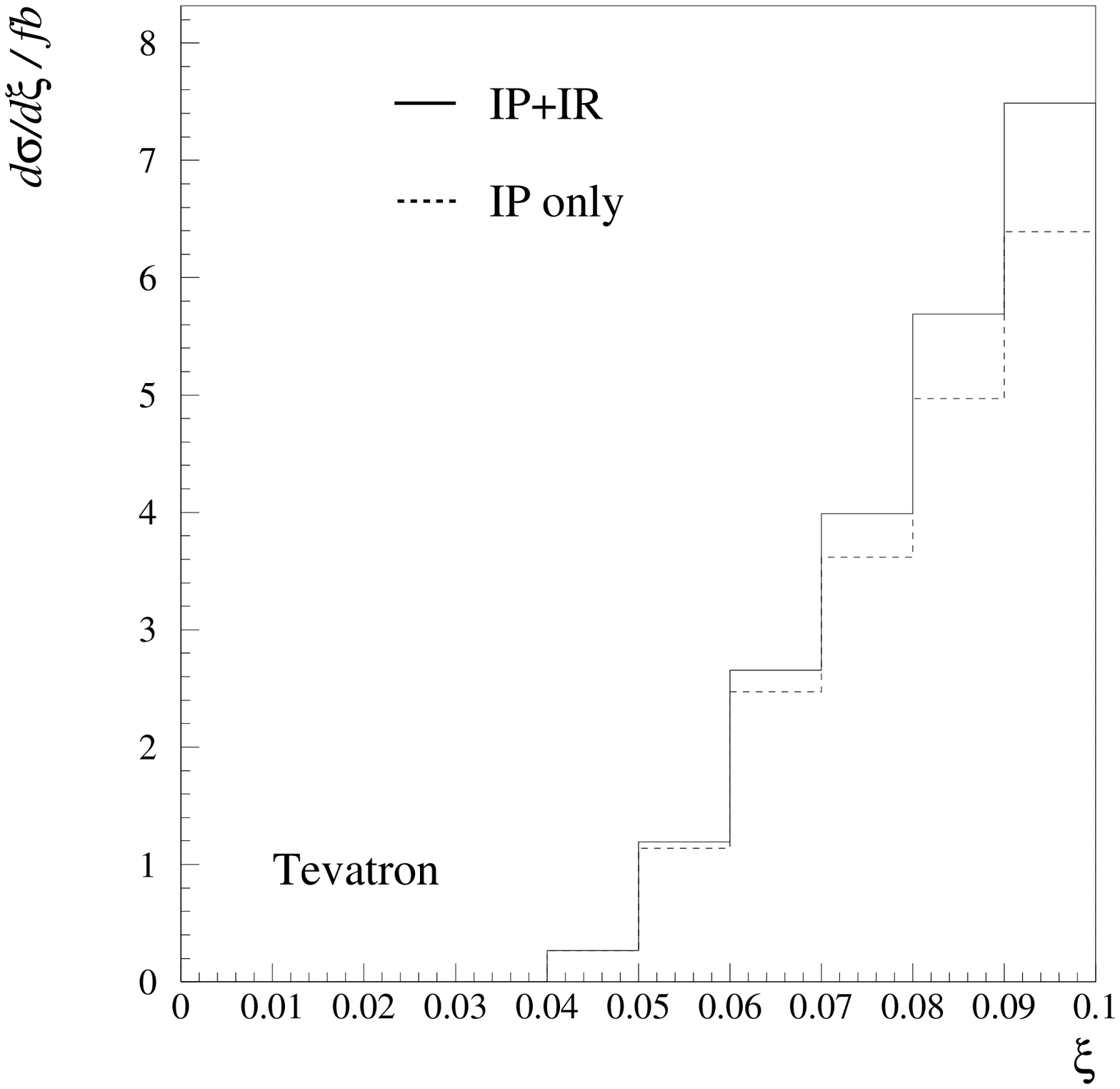,width=.55\textwidth}}
\put( 73,0){\epsfig{file=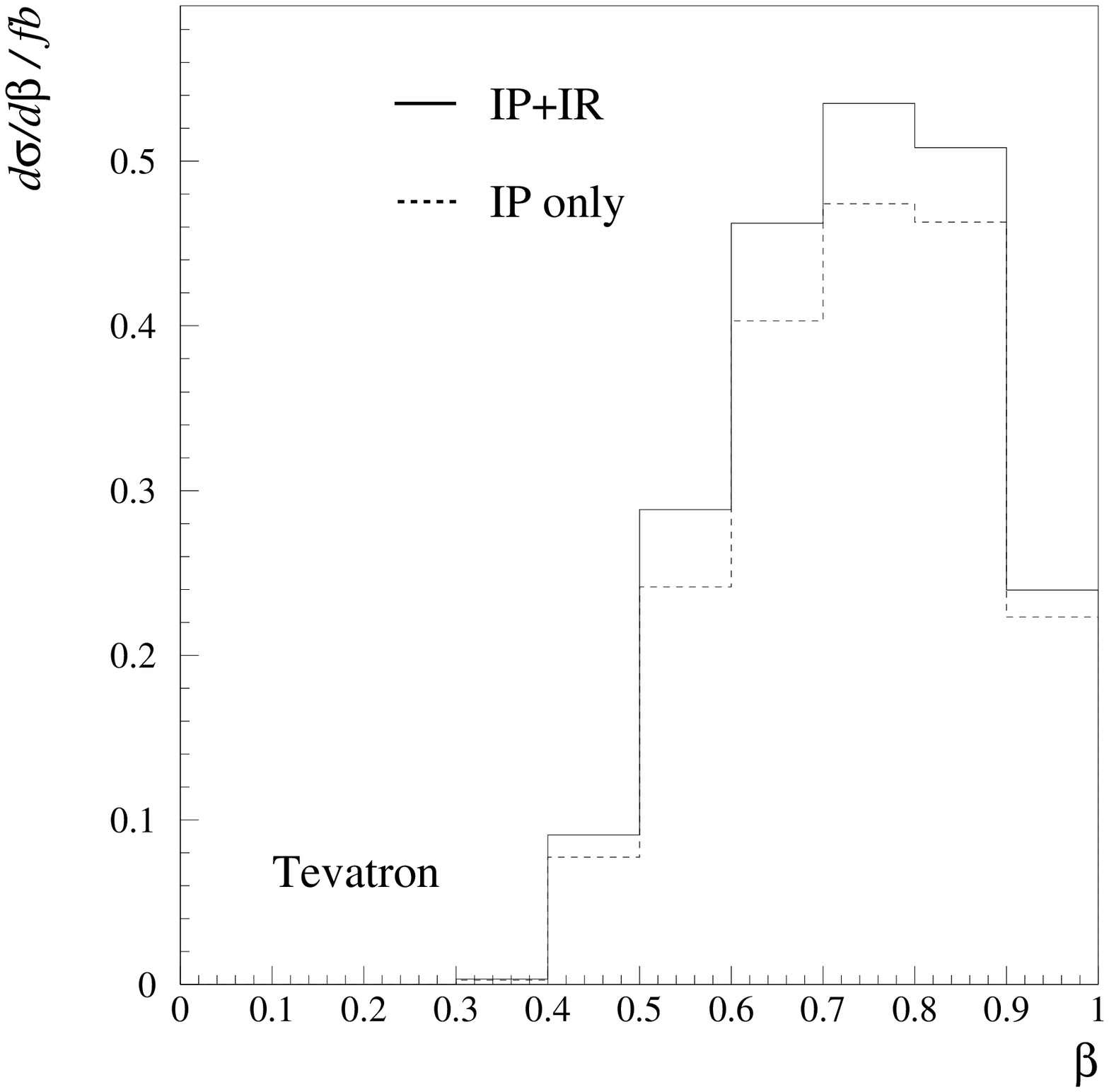,width=.55\textwidth}}
\end{picture}
\caption{In (a) the $\xi$ distribution is shown for double diffractive higgs production with higgs mass $m_h = 115$ GeV at the Tevatron. The $\beta$ distribution is shown in (b).}
\label{xpbh115t}
\end{figure}
\begin{figure}

\begin{picture}(250,85)

\put(0,84){\makebox(0,0){(a)}}
\put(83,84){\makebox(0,0){(b)}}
\put(-10,0){\epsfig{file=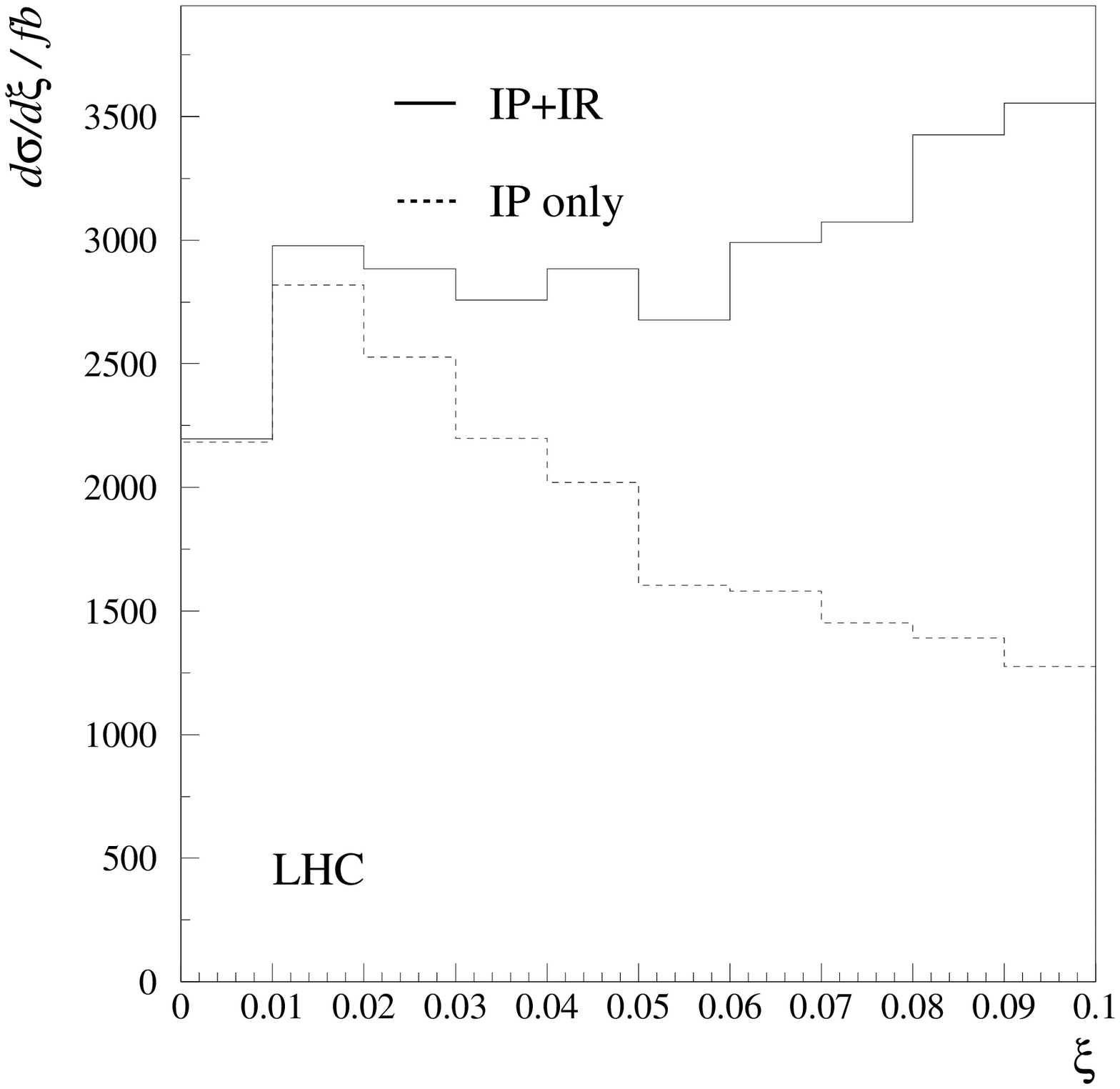,width=.55\textwidth}}
\put( 73,0){\epsfig{file=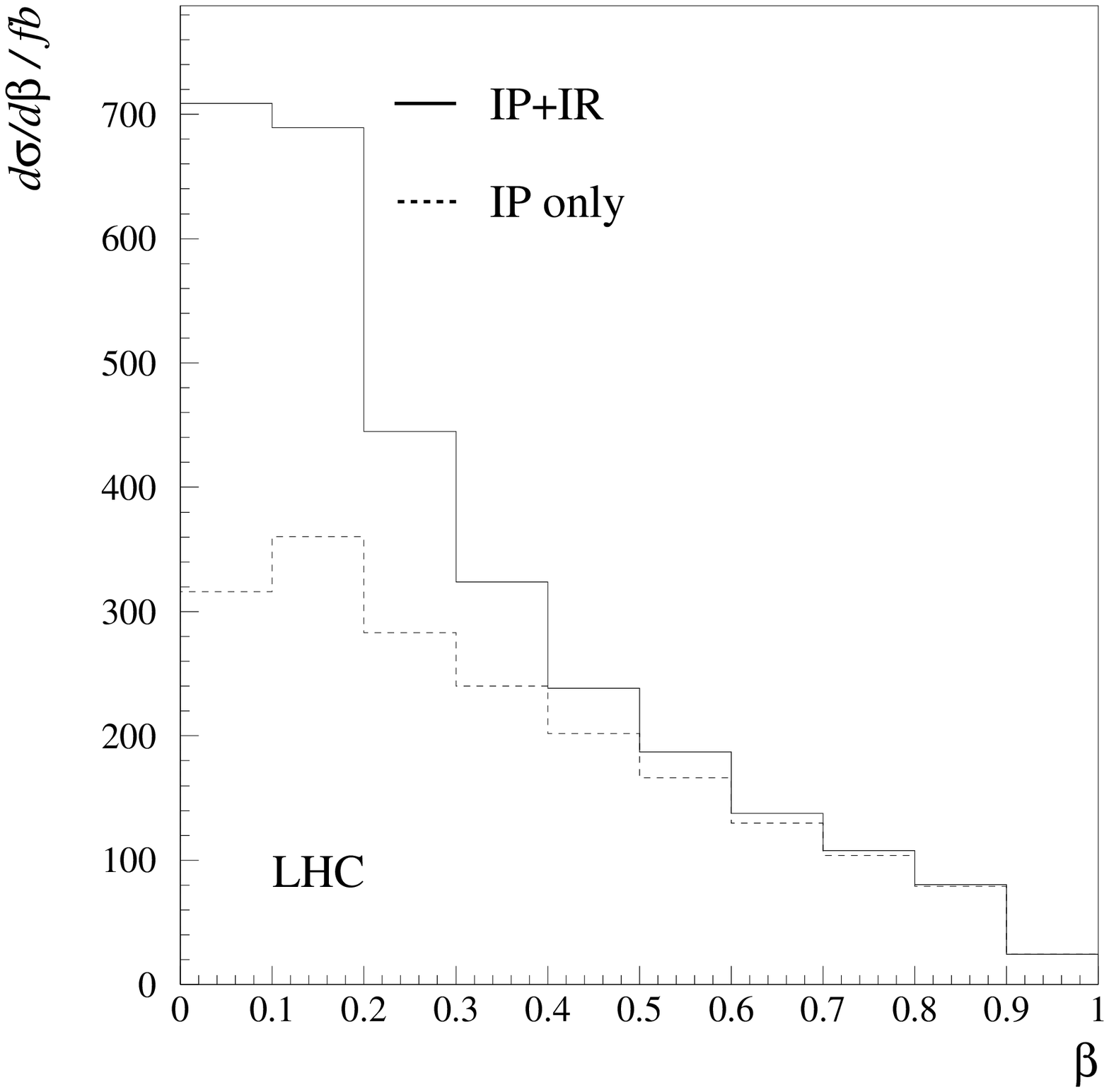,width=.55\textwidth}}
\end{picture}
\caption{In (a) the $\xi$ distribution is shown for double diffractive higgs production with higgs mass $m_h = 115$ GeV at the LHC. The $\beta$ distribution is shown in (b).}
\label{xpbh115l}
\end{figure}

In Figure \ref{hgcs} (a) we show the inclusive higgs production and double diffractive higgs production cross-sections at Tevatron and LHC energies as a function of $m_h$. We require that $\xi < 0.1$ for the diffractive cross-sections. The non-diffractive cross-section was generated using the \herwig 6.1 Monte Carlo and the diffractive cross-section was generated using the \pomwig default parameters for both the $\PO$ and $\RO$ contributions\footnote{Prior to \herwig 6.3, \herwig overestimated the higgs production cross-section by a factor of two. Actually this is quite close to the enhancement typical of NLO QCD corrections \cite{Spira} and so we use the uncorrected results from \herwig 6.1.}. 
The steep fall in the double diffractive cross-section with increasing higgs mass at the Tevatron is due to the limited phase space for high mass resonance production forced by the requirement that $\xi < 0.1$. No such dramatic fall is present at LHC energies. The effect is visible in Figure \ref{hgcs} (b), where we show the ratio of double diffractive to inclusive higgs production.    
The phase space constraints are clearly visible in Figures \ref{xpbh115t} (a) and (b), where we show the $\xi$ and $\beta$ distributions for a higgs mass of $115$ GeV. $\beta$ and $\xi$ are forced to high values since the center of mass energy of the two gluons participating in the higgs production process $\hat s \equiv s \xi{_1} \xi{_2} \beta_1 \beta_2 > m_h^2$. In Figures \ref{xpbh115l} (a) and (b) we show the corresponding distributions at the LHC. Put simply, diffraction as implemented in \pomwig looks very much like a hadron-hadron collision at center of mass energy $s^{'} \le \xi{_1} \xi{_2} s$, which in the case of the Tevatron is of order 200~GeV. It is extremely difficult to produce resonances of 100 GeV or more given this picture. 
\begin{table}[htb]
   \begin{tabular}{|c||c|c|c|c|c|} \hline
      process &  $\sigma / {\rm fb}$   & $\sigma / {\rm fb}$ & $\sigma / {\rm fb}$ & $\sigma / {\rm fb}$ & $\sigma / {\rm fb}$  \\
                & (inclusive) & $\PO$ & $\RO$ & ($\PO + \RO$)      & ($\times S^2$)\\
      \hline  
      $H\rightarrow b\bar{b}$,  $m_h=115$ GeV (Fit 2) & $726.5$ & $ 0.189$ & $0.024$ & $ 0.21$ & $0.02$ \\
      $H\rightarrow b\bar{b}$,  $m_h=115$ GeV (Fit 3) & $726.5$ & $ 0.642$ & $0.024$ & $ 0.67$ & $0.07$ \\
      $H\rightarrow WW$, $m_h=160$ GeV & $260$ & $0.003$ & $0.0002$ & $0.0032$ & \\
      2 $\gamma$,  $E_t>12$ GeV, $\eta<2$ & $72477$ & $265.6$ & $1009.8$ & $1275.4$ & $128$ \\
      2 $\gamma$,  $E_t>20$ GeV, $\eta<1$ & $6345$ & $19.6$ & $63.2$ & $82.8$ & $8.3$ \\
      2 $\gamma$,  $E_t>7$ GeV, (CDF cuts) &  & $684$ & $567$ & $1251$  & $125$ \\
      2 jets $E_t>7$ GeV (CDF cuts)  &  & $243\cdot 10^6$ & $87\cdot 10^6$ & $329.6\cdot 10^6$  & $33\cdot 10^6$  \\
      \hline
    \end{tabular}
\caption{\sl \label{tecatron} Predictions for Tevatron}
\end{table}

\begin{table}[htb]
\begin{center}
   \begin{tabular}{|c||c|c|c|c|} \hline
      process &  $\sigma / {\rm fb}$   & $\sigma / {\rm fb}$ & $\sigma / {\rm fb}$ & $\sigma / {\rm fb}$ \\
                & (inclusive) & $\PO$ & $\RO$ & ($\PO+\RO$)  \\
      \hline  
      $H\rightarrow b\bar{b}$,  $m_h=115$ GeV (Fit 2) & $37660$ & $176$ & $100$ & $276$ \\
      $H\rightarrow b\bar{b}$,  $m_h=115$ GeV (Fit 3) & $37660$ & $254$ & $100$ & $354$ \\
      $H\rightarrow WW$, $m_h=160$ GeV & $19050$ & $97$ & $36$ & $133$  \\
      \hline
    \end{tabular}
\end{center}
\caption{\sl \label{lhc} Predictions for LHC}
\end{table} 
In Table 1 we show the cross-sections for higgs production at the Tevatron for $m_h = 115$ GeV using two different H1 fits for the pomeron structure function, as described in \cite{Adloff:1997sc}. Fit 2 is the \pomwig default, whilst fit 3 has a gluon distribution that peaks at high $\beta$. Both distributions were found to fit the H1 measurement of the diffractive structure function equally well, although measurements of diffractive dijet production at HERA suggest that fit 2 is favoured \cite{Adloff:2001qi}. We also show the cross-section for $m_h =160$ GeV, where the dominant decay mode is to $WW$. 

In order to take gap survival effects into account, we have simulated double diffractive dijet production in the kinematic range measured by the CDF Collaboration \cite{cdfdijet}, namely $0.035 < \xi{_{\bar p}} < 0.095$, $0.01 < \xi{_p} < 0.03$, $-4.2 < \eta_{jets} < 2.4$. The jets were found using a cone algorithm with radius 0.7. CDF measured the cross-section to be $43.6 \pm 4.4~{\rm stat} \pm 21.6~{\rm syst}$ nb, a factor of approximately 10 lower than our result. We therefore estimate that the gap survival factor $S^2 = 0.1$.
This compares favourably with theoretical estimates \cite{Khoze:2000cy, Stelzer, Maor}. Our final cross-section predictions including our estimate of the gap survival factor are shown in the column labelled $ \times S^2$.      

Table 1 also contains our predictions for double diffractive di-photon production for three choices of $E_t^{\gamma}$, $\eta^{\gamma}$ and $\xi$ cuts. Due to the dominance of the quark initiated subprocess, the diphoton production cross-section is dominated by the poorly constrained subleading exchanges at the $\xi$ values which will be available at Run II. We also show the production cross-section in the kinematic range used in the CDF dijet measurement where the subleading component is significantly reduced. Our prediction of 125 events per ${\rm fb}^{-1}$ suggests that this process should be observed in the first year of Run II.     

In Table 2 we show our predictions for LHC energies. In this case, we omit the gap survival factor since we cannot estimate it from data. We note the recent estimate of $S^2 \sim 0.01$ at LHC energies \cite{Khoze:2000cy}.   
\section{Summary}

We have used the \pomwig Monte Carlo generator to estimate the double diffractive higgs boson and di-photon production rates at the Tevatron and LHC. We conclude that the higgs production rate is too small to be observable at the Tevatron. Double diffractive di-photon production, an interesting process in its own right, should be observable within the first year of Tevatron Run II. The situation at the LHC is less clear, primarily due to the uncertainty in the gap survival factors at 14 TeV. If $S^2$ is not much smaller than that at the Tevatron, then we would expect of order 10 events per ${\rm fb}^{-1}$ for $m_h = 115$ GeV, which is a sufficiently high rate to see a signal if the background can be controlled.
        
\section*{Acknowledgements}
We thank Mike Albrow, Valery Khoze, Misha Ryskin and Mike Seymour for
helpful discussions.

\end{document}